\documentclass[%
 reprint,
 amsmath,amssymb,
 aps,
]{revtex4-2}
\usepackage{graphicx}% Include figure files
\usepackage{dcolumn}% Align table columns on decimal point
\usepackage{bm}% bold math
\begin{document}
\preprint{APS/123-QED}
\title{Molecular Hybridization Induced Antidamping and Sizable Enhanced Spin-to-Charge Conversion in Co$_{20}$Fe$_{60}$B$_{20}$/$\beta$-W/C$_{60}$ Heterostructures}% Force line breaks with \\
%\thanks{A footnote to the article title}%
\author{Antarjami Sahoo 1, Aritra Mukhopadhyaya 2, Swayang Priya Mahanta 1, Md. Ehesan Ali 2, Subhankar Bedanta 1,3}
\affiliation{1 Laboratory for Nanomagnetism and Magnetic Materials (LNMM), School of Physical Sciences, National Institute of Science Education and Research (NISER), An OCC of Homi Bhabha National Institute (HBNI), Jatni 752050, Odisha, India}
\affiliation{2 Institute of Nano Science and Technology, Knowledge City, Sector-81, Mohali, Punjab 140306, India}
\affiliation{3 Center for Interdisciplinary Sciences (CIS), National Institute of Science Education and Research (NISER), An OCC of Homi Bhabha National Institute (HBNI), Jatni, Odisha 752050, India}
\author{Md. Ehesan Ali}%
 \email{ehesan.ali@inst.ac.in}
\author{Subhankar Bedanta}%
 \email{sbedanta@niser.ac.in}
\begin{abstract}
Development of power efficient spintronics devices has been the compelling need in the post-CMOS technology era. The effective tunability of spin-orbit-coupling (SOC) in bulk and at the interfaces of hybrid materials stacking is a prerequisite for scaling down the dimension and power consumption of these devices. In this work, we demonstrate the strong chemisorption of C$_{60}$ molecules when grown on the high SOC $\beta$-W layer. The parent CFB/$\beta$-W bilayer exhibits large spin-to-charge interconversion efficiency, which can be ascribed to the interfacial SOC observed at the Ferromagnet/Heavy metal interface. Further, the adsorption of C$_{60}$ molecules on $\beta$-W reduces the effective Gilbert damping by $\sim$15$\%$ in the CFB/$\beta$-W/C$_{60}$ heterostructures. The anti-damping is accompanied by a gigantic $\sim$115$\%$ enhancement in the spin-pumping induced output voltage owing to the molecular hybridization. The non-collinear Density Functional Theory calculations confirm the long-range enhancement of SOC of $\beta$-W upon the chemisorption of C$_{60}$ molecules, which in turn can also enhance the SOC at the CFB/$\beta$-W interface in CFB/$\beta$-W/C$_{60}$ heterostructures. The combined amplification of bulk as well interfacial SOC upon molecular hybridization stabilizes the anti-damping and enhanced spin-to-charge conversion, which can pave the way for the fabrication of power efficient spintronics devices.                     
\end{abstract}
\maketitle
\section{Introduction}
Spintronic logic and memory devices have proven to be one of the most suitable research domains to meet the ultra-low power consumption demand in the post-Complementary Metal Oxide Semiconductor (CMOS) technology era. Especially, with the advent of artificial intelligence and the Internet of Things (IoT), the further scaling down of CMOS technology can reach its physical limits in size, speed, and static energy consumption. The conceptualized spin orbit torque magnetic random access memory (SOT-MRAM) devices which take the advantage of spin Hall effect (SHE) can bring down the energy consumption to femto Joule from the pico Joule scale \cite{guo2021spintronics,ramaswamy2018recent}. The SHE based magnetization switching mechanism in SOT-MRAMs also offers much improved endurance owing to the separation in data writing and reading paths. Though these potentials of SOT-MRAMs have attracted major foundries, several challenges need to be addressed before the commercialization of SOT-MRAMs \cite{guo2021spintronics,ramaswamy2018recent}. The increase of writing efficiency to reduce power consumption is one of those aspects which requires significant consideration. In this context, the spin Hall angle, $\theta_{SH}$ (J$_{S}$ ⁄J$_{C}$ ) of the nonmagnetic layer present in the SOT-MRAMs, where J$_{C}$ and J$_{S}$ are the charge and spin current densities, respectively, plays a critical role in determining the writing efficiency \cite{wang2014determination}. The efficient charge to spin interconversion can lead to the faster switching of magnetization of the adjacent magnetic layer via SHE. Hence, various types of heavy metals (HMs), like Pt, Ta, W, Ir etc. have been investigated in the past two decades to reduce the power consumption of future spintronic devices \cite{fache2020determination,hait2022spin,kim2015dependence}. On a similar note, the Rashba-Edelstein effect (REE) occurring at the interfaces with spatial inversion symmetry breaking and high spin orbit coupling (SOC) has also the potential for the manifestation of efficient charge to spin interconversion\cite{sanchez2013spin,bangar2022large,koo2020rashba}. Hence, the combination of SHE and REE can be the most suitable alternative for the development of power efficient spintronics application.
\par
Among all the heavy metals, highly resistive ($\rho_{\beta-W} \sim 100 - 300 
\mu\Omega$ cm) metastable $\beta$-W possesses the largest $\theta_{SH}$ $\sim$ -0.3 to -0.4 \cite{sui2017giant,mchugh2020impact,demasius2016enhanced,saito2019increase,bai2020simultaneously}, which makes it a strong candidate for SOT-MRAM devices. Usually, additional reactive gases, like O$_2$, N$_2$, and F are employed to stabilize the A15 crystal structure of $\beta$-W \cite{mchugh2020impact} and consequently, a larger $\theta_{SH}$ is realized. For example, Demasius et al., have been able to achieve $\theta_{SH}$ $\sim$ -0.5 by incorporating the oxygen into the tungsten thin films \cite{demasius2016enhanced}. Interface engineering also acts as a powerful tool for enhancing the writing efficiency in $\beta$-W based SOT-MRAM devices \cite{lu2019enhancement,li2019interface,yang2020enhancement}. For instance, the presence of an interfacial atomically thin $\alpha$-W layer in CoFeB/$\alpha$-W/$\beta$-W trilayer suppresses the spin backflow current, resulting in a 45$\%$ increase in the spin mixing conductance \cite{lu2019enhancement}. Further, the REE evolved at the W/Pt interface owing to the charge accumulation generates an additional spin orbit field on the adjacent ferromagnet (FM) NiFe (Py) layer \cite{karube2020anomalous}. The coexistence of SHE and REE has also been reported in CoFeB/$\beta$-Ta and NiFe/Pt bilayers, where the interfacial SOC arising at the FM/HM interface plays a vital role in the spin-to-charge interconversion phenomena \cite{allen2015experimental,hao2022significant}. More interestingly, a recent theoretical work has predicted the interfacial SOC mediated spin Hall angle of Pt can be 25 times larger than the bulk value in NiFe/Pt heterostructure \cite{wang2016giant}. The interfacial SOC mediated spin accumulation has also been reported to occur at the Rashba-like $\beta$-Ta/Py interface without flowing the DC current \cite{behera2016anomalous}. The spin pumping induced by the ferromagnetic resonance results in non-equilibrium spin accumulation at the interface which consequently reduces the effective Gilbert damping of the $\beta$-Ta/Py bilayer. The reduction in effective damping, also termed as antidamping, is similar to the interfacial Rashba like SOT, observed in various HM/FM heterostructures \cite{behera2016anomalous}. The anti-damping phenomena without the requirement of DC current depends on several factors, like SOC of HM, strength of built in electric field at the interface, interface quality etc. Hence, the interface engineering via tuning the interfacial SOC in $\beta$-W based HM/FM heterostructures can be the path forward for developing power efficient SOT-MRAM devices. 
\par
Till the date, most of the interface engineering research have been focused on employing an additional metallic or oxide layer in the HM/FM system for the enhancement of spin-to-charge interconversion efficiency. Whereas, the organic semiconductors (OSCs) can also be incorporated in the HM/FM system to fabricate hybrid power efficient spintronic devices owing to their strong interfacial hybridization and charge transfer nature at metal/OSC interface \cite{pandey2023perspective}. Recently, the SOC of Pt has been found to be enhanced due to the on-surface physical adsorption of C$_{60}$ (fullerene) molecules in YIG/Pt/C$_{60}$ trilayer \cite{alotibi2021enhanced}.  However, the $\theta_{SH}$ of Pt is usually found to be smaller compared to $\beta$-W and it is important to investigate the magnetization dynamics and spin to charge conversion phenomena in FM/$\beta$-W/C$_{60}$ heterostructures. Hence, in this article, we report the effect of molecular hybridization at $\beta$-W/C$_{60}$ interface on magnetization dynamics and spin-to-charge conversion phenomena in Co$_{20}$Fe$_{60}$B$_{20}$ (CFB)/$\beta$-W/C$_{60}$ heterostructures. The molecular hybridization reduces the effective Gilbert damping and also enhances the spin-to-charge conversion efficiency owing to the enhanced SOC of $\beta$-W and consequent strengthening of possible Rashba-like interaction at the CFB/$\beta$-W interface. The strong chemisorption at the $\beta$-W/C$_{60}$ interface and evolution of enhanced SOC of $\beta$-W upon the molecular hybridization have also been confirmed by the first principle density functional theory (DFT) based calculations.
\section{Experimental and Computational Methods}
Four different types of heterostructures with CFB (7 nm)/$\beta$-W (2.5, 5 nm) (Figure \ref{fig:schematics} (a)) and CFB (7 nm)/ $\beta$-W (2.5, 5 nm)/C$_{60}$ (13 nm) (Figure \ref{fig:schematics} (b)) stackings were fabricated on Si/SiO$_{2}$ (300 nm) substrates for the investigation of magnetization dynamics and spin pumping phenomena. In addition, the CFB (7 nm)/$\beta$-W (10, 13 nm) heterostructures were also fabricated to reaffirm the stabilization of $\beta$-W. The heterostructure stackings and their nomenclatures are mentioned in Table \ref{tab:structure}.
\begin{table*}
\caption{\label{tab:structure}Details of the heterostructures and their nomenclatures}
\begin{ruledtabular}
\begin{tabular}{ccc}\\
 Sl. No. & Stacking & Nomenclature\\ \hline
1 & Si/SiO$_{2}$ (300 nm)/CFB (7 nm)/$\beta$-W (2.5 nm) & CFW1\\
2 & Si/SiO$_{2}$ (300 nm)/CFB (7 nm)/$\beta$-W (5 nm) & CFW2\\
3 & Si/SiO$_{2}$ (300 nm)/CFB (7 nm)/$\beta$-W (2.5 nm)/C$_{60}$ (13 nm) &	CFWC1\\
4 & Si/SiO$_{2}$ (300 nm)/CFB (7 nm)/$\beta$-W (5 nm)/C$_{60}$ (13 nm)	& CFWC2\\
\end{tabular}
\end{ruledtabular}
\end{table*}
The CFB and $\beta$-W layers were grown by DC magnetron sputtering, while the Effusion cell equipped in a separate chamber (Manufactured by EXCEL Instruments, India) was used for the growth of the C$_{60}$ over layers in the CFWC series.
While preparing the CFWC1 and CFWC2, the samples were transferred in-situ into the chamber with Effusion cell in a vacuum of $\sim$10$^{-8}$ mbar for the deposition of C$_{60}$. Before the fabrication of heterostructures, thin films of CFB, $\beta$-W, and C$_{60}$ were prepared for thickness calibration and study of magnetic and electrical properties. The base pressure of the sputtering chamber and chamber with Effusion cells were usually maintained at $\sim$ 4$\times$ 10$^{-8}$ mbar and $\sim$ 6$\times$10$^{-9}$ mbar, respectively. The structural characterizations of individual thin films and heterostructures were performed by x-ray diffraction (XRD), x-ray reflectivity (XRR), and Raman spectrometer. 
The magneto-optic Kerr effect (MOKE) based microscope and superconducting quantum interference device based vibrating sample magnetometer (SQUID-VSM) were employed for the static magnetization characterization and magnetic domain imaging. The magnetization dynamics was investigated by a lock-in based ferromagnetic resonance (FMR) spectrometer manufactured by NanOsc, Sweden. The heterostructures were kept in a flip-chip manner on the co-planner waveguide (CPW). The FMR spectra were recorded in the 4-17 GHz range for all the samples. The FMR spectrometer set-up is also equipped with an additional nano voltmeter using which spin-to-charge conversion phenomena of all the devices were measured via inverse spin Hall effect (ISHE) with 5-22 dBm RF power.
The contacts were given at the two opposite ends of 3 mm $\times$ 2 mm devices using silver paste to measure the ISHE induced voltage drop across the samples. The details of the ISHE measurement set-up are mentioned elsewhere \cite{singh2019inverse,gupta2021simultaneous}.
\par
Density functional theory (DFT)-based electronic structure calculations were performed in the Vienna Ab-initio simulation package (VASP) \cite{kresse1996efficient,kresse1999ultrasoft} to understand the interface's chemical bonding and surface reconstructions. The plane wave basis sets expand the valance electronic states, and the core electrons are treated with the pseudopotentials. The core-valance interactions are considered with the Projected Augmented Wave method.  The exchange-correlation potentials are treated with Perdew, Bruke and Ernzerof (PBE) \cite{perdew1996generalized} functional which inherits the Generalized Gradient Approximation (GGA). This functional produces a reliable understanding of similar kinds of interfaces. The convergences in the self-consistent field iterations were ensured with a plane-wave cutoff energy of 500 eV and a tolerance of 10$^{-6}$ eV/cycle. 
A D3 dispersion correction term, devised by Grimme, accounts for the long-range interaction terms was employed in the calculations. The optimized unit cell parameter obtained from the aforementioned methods for the cubic A15 crystal of the $\beta$-W is 5.014 \AA, which resembles the experimental parameter of 5.036 \AA. 
A 5$\times$2$\times$1 repetition is used to construct the (210) surface unit cell of the $\beta$-W to model the surface supercell. The lower two atomic layers were fixed at the bulk, and the remaining three layers were allowed to relax during the geometry optimization. The surface layer of $\beta$-W contains the C$_{60}$ molecules. To understand the effect of the spin-orbit coupling interactions, we have performed the non-collinear DFT calculations as implemented in VASP. The E$_{SOC}$ calculated from these calculations quantifies the strength of the SOC term in the Hamiltonian.
\section{Results and Discussion}
\begin{figure*}
\includegraphics[scale=0.6]{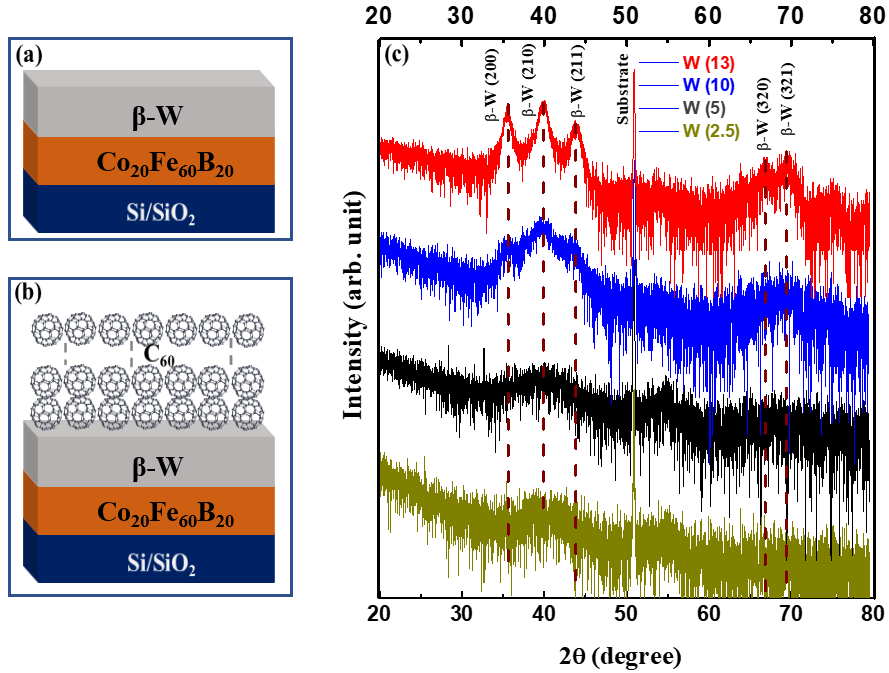}
\caption{\label{fig:schematics}Schematics of (a) Si/SiO$_2$/CFB/$\beta$-W and (b) Si/SiO$_2$/CFB/$\beta$-W/C$_{60}$ heterostructures, (c) GIXRD patterns of Si/SiO$_2$/CFB/$\beta$-W heterostructures with various thicknesses of $\beta$-W.}
\end{figure*}
The grazing incidence x-ray diffraction (GIXRD) was performed for all the heterostructures. The XRD patterns of CFB/$\beta$-W heterostructures with different thicknesses of $\beta$-W are shown in Figure \ref{fig:schematics} (c). The presence of (200), (210) and (211) Bragg’s peaks of W at 35.5$^\circ$, 39.8$^\circ$ and 43.5$^\circ$ indicate the stabilization of metastable $\beta$ phase of W (A-15 crystal structure) \cite{hait2022spin,behera2017capping}. In addition, we have also observed the (320) and (321) Bragg’s peaks of W, which further suggests the growth of polycrystalline $\beta$-W. The relative intensity of (320) and (321) Bragg’s peaks of W is lower compared to (200), (210) and (211) Bragg’s peaks, consistent with previous reports \cite{behera2017capping}. The Bragg’s peaks are more prominent for heterostructures with thicker W layers as diffraction intensity increases with the increase in W thickness. The XRD patterns for CFWC1 and CFWC2 are similar to that for CFW1 and CFW2, respectively, as the thickness of $\beta$-W are same.  Here, we have not used the reactive gases like, O$_2$ and N$_2$ for the growth of $\beta$-W unlike some previous report \cite{mchugh2020impact}. The resistivity of W films with thicknesses 2.5, 5, 10 nm were measured by standard four probe methods. The resistivity decreases with increase in thickness of W and were found in between $\sim$ 300-100 $\mu \Omega$-cm, further confirming the growth of $\beta$ phase of W \cite{hait2022spin,behera2017capping}. We do not also observe the (110), (200), (210) Bragg’s peaks for the bcc $\alpha$-W in the XRD patterns and the $\alpha$-W would have also exhibited one order less resistivity compared to what we have observed \cite{behera2017capping}. The stabilization of pure $\beta$ phase of W is quite important for future SOT device fabrication and hence, we can expect a high spin-to-charge conversion efficiency in our CFB/$\beta$-W heterostructures owing to high SOC of $\beta$-W \cite{sui2017giant,behera2017capping}.
\par
The XRR measurements were performed for all the samples in both the CFW and CFWC series to confirm the desired thickness of individual layers and to investigate the interface quality. Figure S1 (Supporting Information) shows the XRR patterns of all the heterostructures considered for the present study. The experimental data were fitted using GenX software and the simulated patterns are also shown in Figure S1 (red curves). The presence of Kiessig oscillations for all the films infer the absence of a high degree of interfacial disorder and dislocations. The relative peak positions and intensity of the simulated patterns agree quite well with the experimentally observed low angel XRR data. The fit provides the anticipated thickness of individual layer in each sample as mentioned in Table \ref{tab:structure}. The interface roughness for all the heterostructures were found in between 0.2-0.5 nm, further inferring the high quality growth of both the series of samples. Figure S2 (Supporting Information) displays the Raman spectra of 13 nm C$_{60}$ film grown on Si/SiO$_2$ (300 nm) substrate with the same growth condition as in the heterostructures. The presence of A$_{g}$(2) and H$_{g}$(8) Raman modes of C$_{60}$ around $\sim$ 1460 cm$^{-1}$ and 1566 cm$^{-1}$, respectively confirms the growth of C$_{60}$ film \cite{meilunas1991infrared,cataldo2000raman}. In addition, the Raman mode around $\sim$ 495 cm$^{-1}$ corresponding to A$_{g}$(1) mode of C$_{60}$ is also observed in the Raman spectrum. The anticipated thickness of C$_{60}$ in the C$_{60}$ thin film, CFWC1 and CFWC2 has also been confirmed from the XRR measurements. The Raman spectrum of our C$_{60}$ film grown by effusion cells are quite similar to those prepared by different solution methods in HCl or N$_2$ atmosphere \cite{meilunas1991infrared,cataldo2000raman}. The saturation magnetization and the magnetic domain images of all the heterostructures are found to be similar (see Supporting Information) as the bottom CFB layer is same for all the heterostructures.\par
The magnetization relaxation and propagation of spin angular momentum in the CFB thin film and the heterostructures in both the CFW and CFWC series were studied to explore the effect of high resistive $\beta$-W and $\beta$-W/C$_{60}$ bilayer by in-plane FMR technique. The heterostructures are placed in a flip-chip manner on CPW as shown in the schematics in Figure S4 (a) (Supporting Information).  Figure S4 (c) shows the typical FMR spectra of CFW1 and CFWC1 heterostructures measured in the 4-17 GHz range. All the FMR spectra were fitted to the derivative of symmetric and antisymmetric Lorentzian function to evaluate the resonance field ($H_{res}$) and linewidth ($\Delta H$) \cite{singh2020large}: 
\begin{widetext}
\begin{equation}
    FMR Signal = K_1 \frac{4(\Delta H)(H-H_{res})}{[(\Delta H)^2 + 4 (H - H_{res})^2]^2} - K_2 \frac{(\Delta H)^2 - 4(H-H_{res})^2}{[(\Delta H)^2 + 4 (H - H_{res})^2]^2} + Offset,
    \label{eq:FMR}
\end{equation}
\end{widetext}
where K$_1$ and K$_2$ are the antisymmetric and symmetric absorption coefficients, respectively.
\begin{figure*}
\includegraphics[scale=0.6]{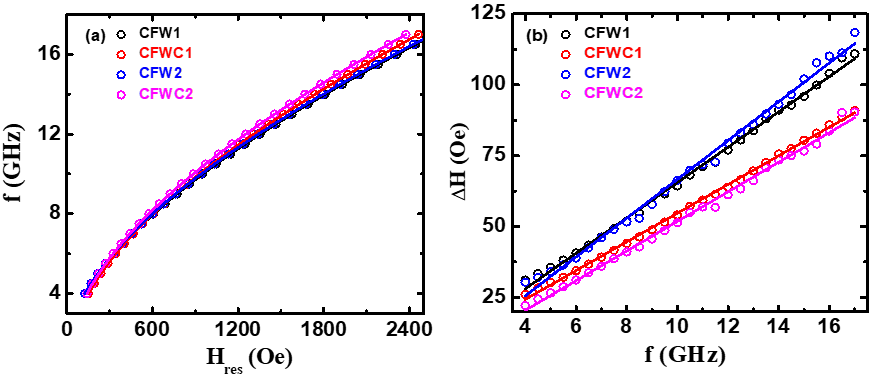}
\caption{\label{fig:FMRFig}(a) Frequency ($f$) versus resonance field ($H_{res}$) and (b) linewidth ($\Delta H$) versus frequency ($f$) behaviour for various heterostructures. The solid lines are the best fits to equation \ref{eq:Kittel} and \ref{eq:damping}.}
\end{figure*}
The extracted $H_{res}$  and $\Delta H$ values at different resonance frequencies ($f$) of all the heterostructures are shown in Figure \ref{fig:FMRFig} (a-b). The $f$ vs $H_{res}$ of different samples in the CFW and CFWC series are plotted in Figure \ref{fig:FMRFig} (a). The $f$ vs $H_{res}$ plots are fitted by using equation \ref{eq:Kittel}  \cite{singh2020large}:
\begin{equation}
    f = \frac{\gamma}{2 \pi} \sqrt{(H_K + H_{res})(H_K +H_{res}+4\pi M_{eff})},
    \label{eq:Kittel}
\end{equation}
where$$4\pi M_{eff} = 4\pi M_S + \frac{2K_S}{M_S t_{FM}}$$
and H$_K$, K$_S$, and t$_{FM}$ are the anisotropy field, perpendicular surface anisotropy constant, and the thickness of FM, respectively. Here, $\gamma$ is the gyromagnetic ratio and $4 \pi M_{eff}$ represents the effective magnetization. The $4 \pi M_{eff}$ extracted from the fitting gives similar values as compared with the saturation magnetization value ($4 \pi M_{S}$) calculated from the SQUID-VSM. Further, the effective Gilbert damping constant ($\alpha_{eff}$) and hence, the magnetization relaxation mechanism are studied from the resonance frequency dependent FMR linewidth behavior. The $\Delta H$ vs $f$  plots are shown in Figure \ref{fig:FMRFig} (b). The linear dependency of $\Delta H$ on $f$ indicates the magnetic damping is mainly governed by intrinsic mechanism via electron-magnon scattering rather than the extrinsic two magnon scattering. The $\Delta H$ vs $f$ plots are fitted by the following linear equation \cite{singh2020large} to evaluate  the $\alpha_{eff}$.
\begin{equation}
    \Delta H = \Delta H_0 + \frac{4 \pi \alpha_{eff}}{\gamma}f,
    \label{eq:damping}
\end{equation}
where the  $\Delta H_0$ is the inhomogeneous linewidth broadening. The $\alpha_{eff}$ values for all the heterostructures and CFB thin film obtained from the fitting are shown in Table \ref{tab:magdata}. The $\alpha_{eff}$  value for CFW series ($\sim$0.0075$\pm$0.0001 for CFW1 and $\sim$0.0080$\pm$0.0001 for CFW2) are found to be larger compared to that of the CFB thin film ($\sim$0.0059$\pm$0.0001). The enhancement of $\alpha_{eff}$ indicates the possible evolution of spin pumping mechanism in the CFB/$\beta$-W bilayers. Interestingly, the $\alpha_{eff}$ decreases to $\sim$ 0.0065$\pm$0.0001 upon the deposition of C$_{60}$ molecules on CFB/$\beta$-W bilayers in CFWC series. The significant change in $\alpha_{eff}$  for the CFB/$\beta$-W/C$_{60}$ heterostructures compared to CFB/$\beta$-W bilayers infers the modification of physical properties of $\beta$-W layer in CFB/$\beta$-W/C$_{60}$. The deposition of C$_{60}$ molecules can lead to the metal/molecule hybridization at the $\beta$-W/C$_{60}$ interface, which in turn can alter the properties of $\beta$-W.  \par
\begin{figure*}
\includegraphics[scale=1.2]{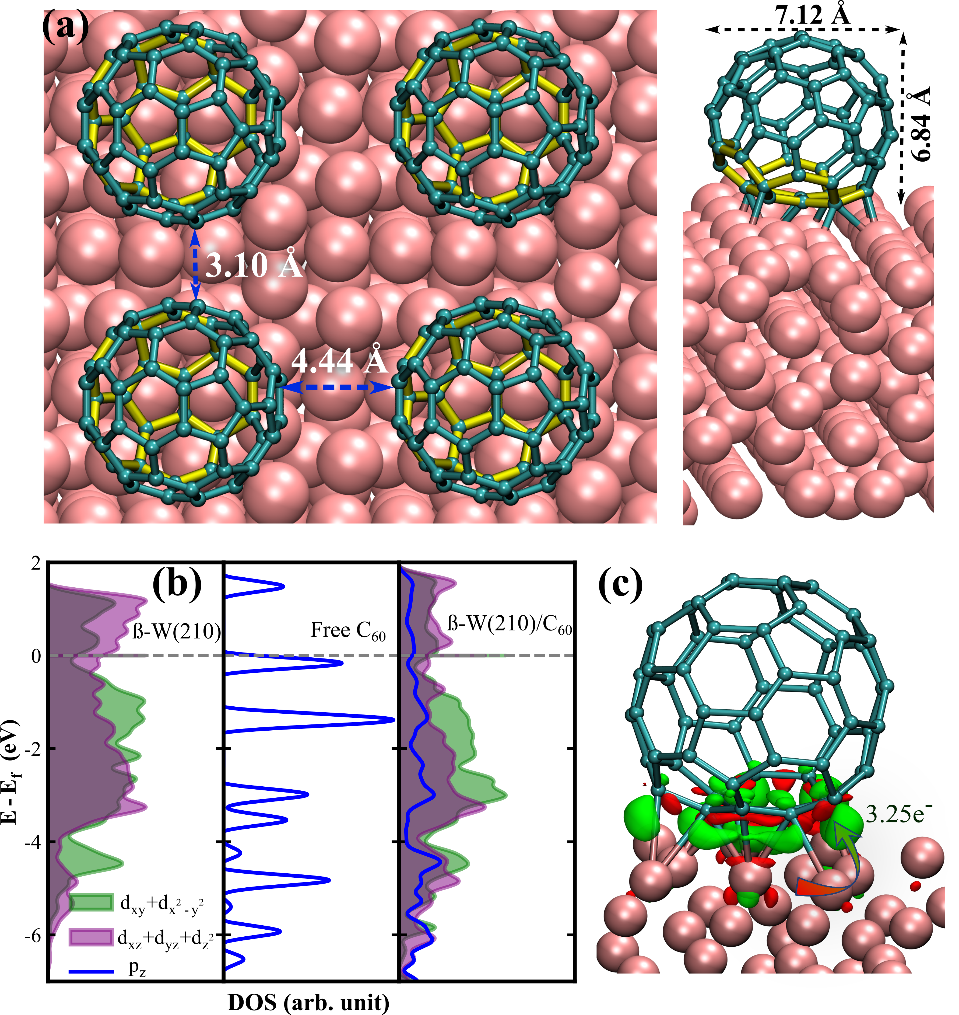}
\caption{\label{fig:DOS}(a) The extended simulation supercell for the C$_{60}$ on $\beta$-W(210) substrate. The left panel shows the top view of the surface supercell (along the z-axis), and the right panel shows the side view of the same. The pink balls of larger size and cyan balls of smaller size represent the tungsten and carbon atoms, respectively. The yellow bonds highlight the part of the C$_{60}$ which takes part in the interface formation. The double-headed dotted arrows quantify the diameter of the C$_{60}$ spheres in two directions. (b-c) The modification in the electronic structure due to chemisorption of the C$_{60}$ molecule on the $\beta$-W. (b) The atom projected orbital resolved partial density of states of $\beta$-W(210), C$_{60}$, and $\beta$-W(210)/C$_{60}$, and (c) The electron density redistribution due to chemisorption. The red and green iso-surfaces depict electron density depletion and accumulation of the electron density at the interface, respectively. The bi-coloured arrow depicts the direction of the electron transfer process.}
\end{figure*}

The DFT based first principle calculations were performed to elucidate further the molecular hybridization at the $\beta$-W/C$_{60}$ interface and its consequences on the magnetization dynamics of CFB/$\beta$-W/C$_{60}$ heterostructures. The extended simulation supercell for the C$_{60}$ on $\beta$-W(210) are shown in Figure \ref{fig:DOS} (a). The C$_{60}$ molecule is observed as strongly chemisorbed onto the $\beta$-W (210) surface with an adsorption energy of -253.5 kcal/mol. The adsorption energy is quite high as compared to the other substrates. For example, the adsorption energy for Co/C$_{60}$ was found to be -90 kcal/mol \cite{sharangi2022effect} while for the Pt/C$_{60}$ interface it is reported to be -115 kcal/mol \cite{shi2011c}. The chemisorption in case of $\beta$-W/C$_{60}$ is quite strong and induces distortion to the spherical shape of the adsorbed C$_{60}$. The distance between two carbon atoms from two opposite hexagons of adsorbed C$_{60}$ is shorter along one direction compared to the other measured in the plane (left panel of Figure \ref{fig:DOS} (a)). The diameter of C$_{60}$ molecules decreases by 0.3 \AA when it is measured perpendicular to the $\beta$-W (210) surface (right panel of Figure \ref{fig:DOS} (a)). 
This distortion can be attributed to the W-C bond formation due to the strong chemisorption at the $\beta$-W/C$_{60}$ interface. This chemisorption strongly alters the electronic structure of the $\beta$-W and C$_{60}$ molecule (Figure \ref{fig:DOS} (b)). The $p_z$ orbital, which accommodates the $\pi$-electrons of the C$_{60}$, hybridizes with the $d$-orbitals of the $\beta$-W atom and forms the hybridised interfacial states. The out-of-plane d-orbitals ($d_{xz}$,$d_{yz}$ and $d_{z^2}$ orbitals) are strongly hybridized with the $p_z$ orbital of the carbon atom over a large energy window near the Fermi energy level (Figure \ref{fig:DOS} and Figure S5 (Supporting Information)). 
The sharp peaks observed in the DOS of free C$_{60}$ layer gets significantly broadened, flattened, and shifted for $\beta$-W/C$_{60}$ stacking. The strong metallo-organic hybridization also modifies the PDOS of various d-orbitals of $\beta$-W. The various d-orbitals become flattened and spread over larger energy spectrum around the Fermi level upon molecular hybridization. The formation of the W-C bond also costs a transfer of 3.25e$^-$ from the interfacial layer of the $\beta$-W to C$_{60}$ molecule (Figure \ref{fig:DOS} (c)). This is relatively higher compared to the previously reported the 0.25e$^-$ transfer from Pt (111) and 3e$^-$ transfer from Cu (111) to the adjacent C$_{60}$ molecule, inferring the metallo-organic hybridization is quite stronger in case of $\beta$-W/C$_{60}$ interface \cite{shi2011c}.Hence, the molecular hybridization of $\beta$-W is expected to alter its physical properties with greater effect and can be considered as an important tool to optimize the spintronics device performances. 
\par
\begin{figure*}
\includegraphics[scale=1]{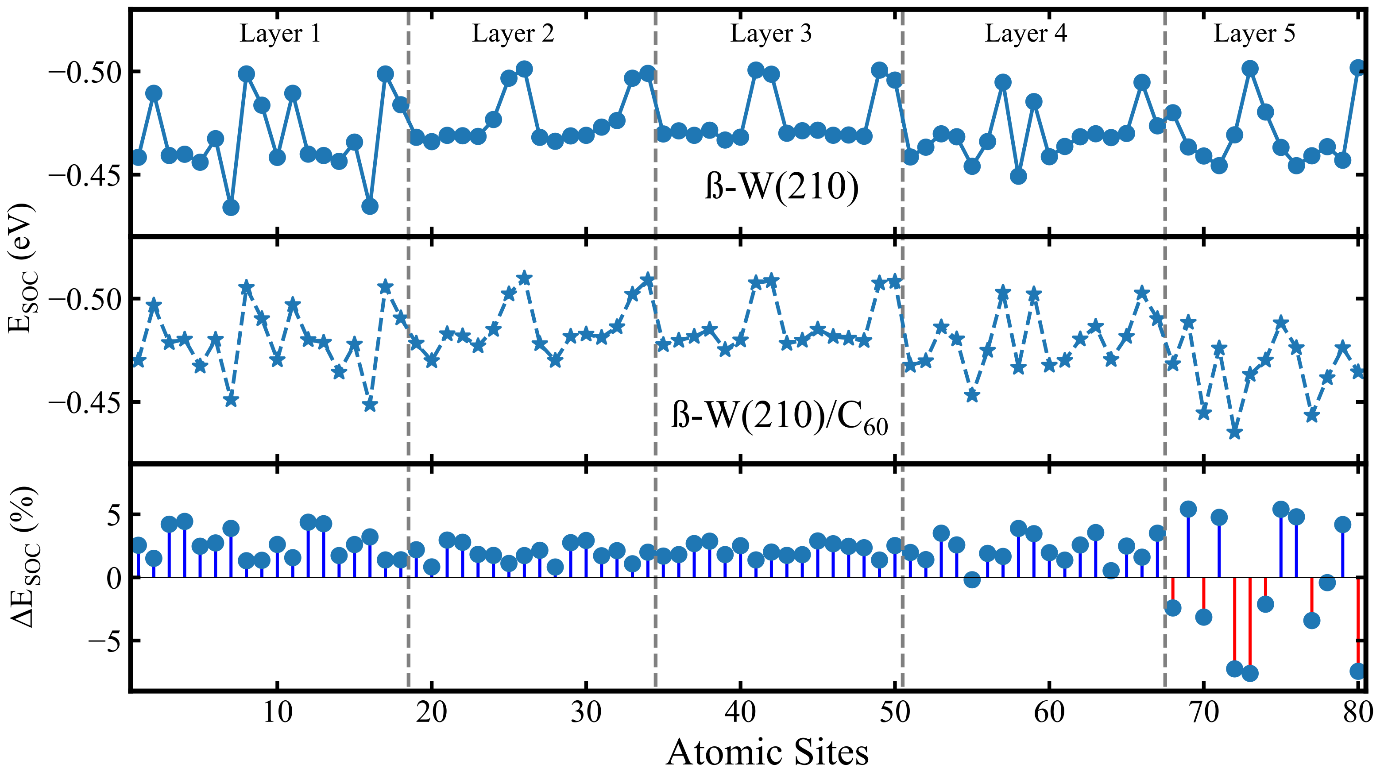}
\caption{\label{fig:SOC}The effect of the chemisorption of the C$_{60}$ molecule at the $\beta$-W(210) surface on the  E$_{SOC}$ of various atomic sites. The percentage change in the E$_{SOC}$ ($\Delta$E$_{SOC}$) is calculated in terms of the change in the E$_{SOC}$ of the bare $\beta$-W(210) substrate. Layer 5 is the interfacial layer that interacts with the C$_{60}$, and layer 1 is the opposite to the $\beta$-W/C$_{60}$ interface layer.}
\end{figure*}

The modified electronic structure was found to carry a long-range effect on the strength of the spin-orbit coupling. The E$_{SOC}$ of bare 2.5 nm $\beta$-W and 2.5 nm $\beta$-W covered with C$_{60}$ molecules, and the variation of the E$_{SOC}$ ($\Delta$E$_{SOC}$) due to $\beta$-W/C$_{60}$ hybridization are shown in Figure \ref{fig:SOC}. The interfacial W atoms involved in the hybridization with C$_{60}$ show a decrease in the E$_{SOC}$. The rest of the W atoms from the surface layer exhibit an increase in the E$_{SOC}$. The lower atomic layers of W also show an increment in the E$_{SOC}$. The W layer, farthest from the $\beta$-W/C$_{60}$ interface (nearer to the CFB/$\beta$-W interface), exhibits the most increased E$_{SOC}$. Hence, the hybridization at the $\beta$-W/C$_{60}$ interface increases the overall spin-orbit coupling strength of the $\beta$-W layer. More importantly, the SOC at the CFB/$\beta$-W interface is enhanced for CFB/$\beta$-W/C$_{60}$ stacking compared to the CFB/$\beta$-W bilayer. The enhanced bulk SOC of $\beta$-W and the interfacial SOC at CFB/$\beta$-W interface can facilitate an efficient spin to charge conversion in CFB/$\beta$-W/C$_{60}$ heterostructures.  
\par
The decrease in damping, usually know as anti-damping, has been observed previously in FM/HM bilayers \cite{behera2016anomalous,gupta2021simultaneous,behera2017capping}. In those systems, the effective damping values become lower than the $\alpha_{eff}$  of the FM layer and this phenomenon has been attributed to the formation of Rashba like interfacial states \cite{behera2016anomalous,behera2017capping}. Similar type of evolution of Rashba like states at the CFB/$\beta$-W interface can be expected due to structural inversion asymmetry and large SOC of $\beta$-W. The spin accumulation at the CFB/$\beta$-W interface can lead to evolution of the non-equilibrium spin states. The non-equilibrium spin states along with the enhanced SOC at CFB/$\beta$-W interface due to molecular hybridization as confirmed from the DFT calculations can generate an additional charge current due to IREE and can also induce the antidamping torque on the magnetization of FM layer. 
The antidamping torque can make the magnetization precession relatively slower and thus decreasing the $\alpha_{eff}$ of the CFB/$\beta$-W/C$_{60}$ heterostructures compared to the CFB/$\beta$-W bilayer. The control of Gilbert damping of FMs by interfacing with adjacent non-magnetic metal/organic bilayers can also provide an alternative to the search for low damping magnetic materials. Especially, the low cost and abundant availability of carbon based organic molecules can be commercially beneficial in optimizing the magnetic damping for spintronic applications. Further, the Gilbert damping modulation can also control the effective spin mixing conductance ($g_{eff}^{(\uparrow \downarrow)}$)  of the heterostructures which also plays a vital role for efficient spin current transport across the interface. Hence, the $g_{eff}^{(\uparrow \downarrow)}$ of all the heterostructures was calculated from the damping constant measurement by equation \ref{geff} \cite{singh2020large}:
\begin{equation}
    g_{eff}^{(\uparrow \downarrow)} = \frac{4 \pi M_s t_{CFB}}{g \mu_B} (\alpha_{CFB/NM} - \alpha_{CFB}),
    \label{geff}
\end{equation}
where g, $\mu_B$ and $t_{CFB}$ are the Landé g factor (2.1), Bohr’s magnetron, and thickness of CFB layer, respectively.   $\alpha_{CFB/NM}$ is the damping constant of bilayer or tri-layers and $\alpha_{CFB}$  is the damping constant of the reference CFB thin film. The $g_{eff}^{(\uparrow \downarrow)}$ for CFW1 and CFW2 (Table \ref{tab:magdata}) are relatively higher compared to the previous reports on FM/$\beta$-W bilayers. Especially, the $g_{eff}^{(\uparrow \downarrow)}$ of CFW2 is one order higher than that reported for Py/$\beta$-W bilayer (1.63$\times$ 10$^{18}$ m$^{-2}$) \cite{behera2017capping}, and 2 order higher compared to that of the YIG/$\beta$-W (5.98 $\times$ 10$^{17}$ m$^{-2}$) \cite{bai2020simultaneously}. This indicates the absence of any significant amount of spin back flow from $\beta$-W layer and high SOC strength of parent $\beta$-W layer in our system. However, the $g_{eff}^{(\uparrow \downarrow)}$  values decrease for the CFWC1 and CFWC2 tri-layers owing to anti-damping phenomena.\par
\begin{figure*}
\includegraphics[scale=0.8]{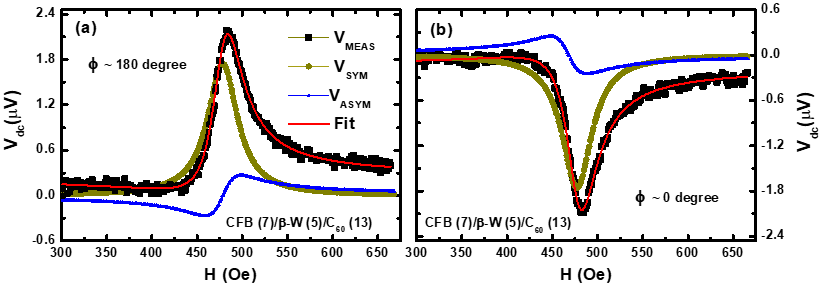}
\caption{\label{fig:ISHEplot1}$V_{MEAS}$, $V_{SYM}$ and $V_{ASYM}$ versus $H$ for CFB (7)/$\beta$-W(5)/C$_{60}$ (13) [CFWC2] heterostructure with $\phi$ $\sim$(a) 180$^\circ$ and (b) 0$^\circ$ measured at 15 dBm RF power. The red curve is Lorentzian fit with equation \ref{eq:ISHE} to $V_{dc}$ vs $H$ plot.}
\end{figure*}

The ISHE measurements were performed for all the heterostructures in CFW and CFWC series to gain more insights about the effect of molecular hybridization in CFB/$\beta$-W/C$_{60}$ on the magnetization dynamics and spin to charge conversion efficiency. Figure \ref{fig:ISHEplot1} shows the typical field dependent DC voltage ($V_{dc}$) measured across the CFB (7 nm)/$\beta$-W (5 nm)/C$_{60}$ (13 nm) heterostructure under FMR conditions. In order to separate the symmetric ($V_{SYM}$) and asymmetric ($V_{ASYM}$) components, the $V_{dc}$ vs $H$ plots were fitted with the following Lorentzian function:
\begin{widetext}
\begin{equation}
    V_{dc} = V_{SYM} \frac{(\Delta H)^2}{(\Delta H)^2 + (H - H_{res})^2} + V_{ASYM} \frac{(\Delta H) (H-H_{res})}{(\Delta H)^2 + (H - H_{res})^2} 
    \label{eq:ISHE}
\end{equation}
\end{widetext}
The extracted field dependent $V_{SYM}$ and $V_{ASYM}$ are also plotted in Figure \ref{fig:ISHEplot1}. Similar type of field dependent $V_{MEAS}$, $V_{SYM}$, and $V_{ASYM}$ are also observed for other samples in both CFW and CFWC series. The $V_{SYM}$ is mainly contributed by the spin pumping voltage ($V_{ISHE}$) and the spin rectification effects arising from the anisotropic magnetoresistance (AMR) [$V_{AMR}$] \cite{singh2020large}.  Whereas, the asymmetric component of the measured voltage arises solely due to anomalous Hall effect and AMR \cite{singh2020large}. The sign of $V_{SYM}$ is reversed when $\phi$ (angel between the perpendicular direction to the applied magnetic field (\textit{H}) and direction of voltage measurement) is changed from 0$^\circ$ to 180$^\circ$ (Figure \ref{fig:ISHEplot1}), confirming the presence of ISHE in our heterostructures. 
\begin{figure*}
\includegraphics[scale=0.7]{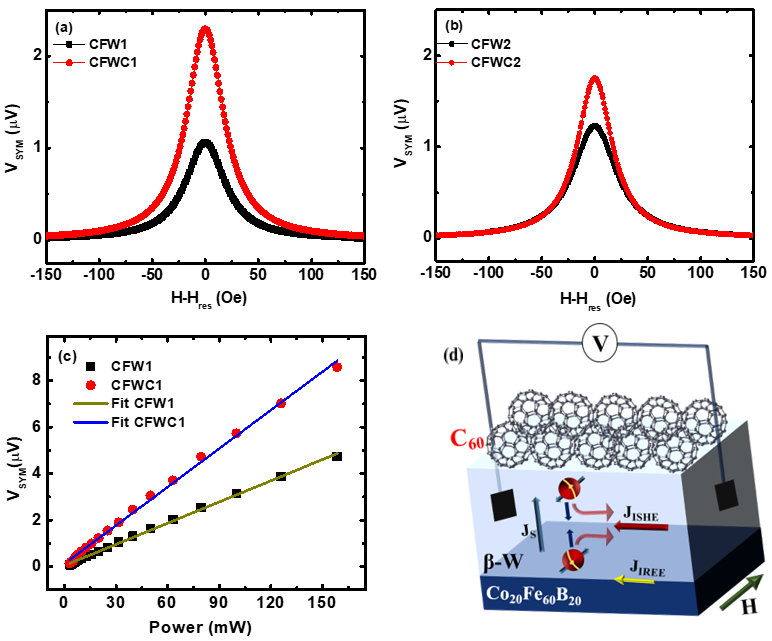}
\caption{\label{fig:ISHEplot2}$V_{SYM}$ versus applied magnetic field with $\phi$ $\sim$ 180$^\circ$  for (a) CFB (7)/$\beta$-W(2.5) [CFW1] and CFB (7)/$\beta$-W(2.5)/C$_{60}$ (13) [CFWC1] and (b) CFB (7)/$\beta$-W(5) [CFW2] and CFB (7)/$\beta$-W(5)/C$_{60}$ (13)  [CFWC2] heterostructures measured at 15 dBm RF power, (c) Power dependent $V_{SYM}$ for CFW1 and CFWC1 (The solid line is the linear fit), (d) Schematic showing the spin-to-charge conversion phenomena in CFB/$\beta$-W/C$_{60}$ heterostructures.}
\end{figure*}
The field dependent $V_{SYM}$ for all the four heterostructures are plotted in Figure \ref{fig:ISHEplot2} (a-b). Interestingly, the $V_{SYM}$ value at the resonance field for CFB/$\beta$-W/C$_{60}$ trilayers is found to be increased compared to that for CFB/$\beta$-W bilayers. The increment is $\sim$ 115$\%$ for $\beta$-W thickness 2.5 nm, while it becomes $\sim$ 20$\%$ for $\beta$-W thickness 5 nm. The gigantic enhancement of $V_{SYM}$ for CFB (7)/$\beta$-W(2.5)/C$_{60}$ (13) infers the modification of SOC of $\beta$-W when capped with organic C$_{60}$ molecules and the presence of an additional spin to charge conversion effect in the heterostructures. 
\begin{table}
\caption{\label{tab:magdata}Effective Gilbert damping, spin mixing conductance, and symmetric component of measured DC voltage for different heterostructures.}
\begin{ruledtabular}
\begin{tabular}{l|c|c|r}
Heterostructures & $\alpha_{eff}$($\pm$0.0001) & $g_{eff}^{(\uparrow \downarrow)}$ (10$^{19} m^{-2}$) & $V_{SYM} (\mu V)$\\
\hline
CFB & 0.0059 &	- & -\\
CFW1 & 0.0075	& 0.87 & 1.08\\
CFW2 & 0.0080 &	1.13 & 1.25\\
CFWC1 & 0.0064	& 0.27 & 2.32\\
CFWC2 & 0.0065	& 0.32 & 1.78\\
\end{tabular}
\end{ruledtabular}
\end{table}
The power dependent spin-to-charge conversion measurements were also performed to further confirm the enhancement of $V_{SYM}$. The spin pumping induced voltage increases linearly with the RF power as shown in Figure \ref{fig:ISHEplot2} (c) for both CFW1 and CFWC1. The $V_{SYM}$ at different RF power is found to be increased for CFWC1 compared to CFW1, which further confirms the  molecular hybridization induced enhanced spin-to-charge conversion. As the thickness, magnetic properties of bottom CFB layer is same for all the heterostructures, the contribution of $V_{AMR}$ is expected to be same for CFB (7 nm)/$\beta$-W(2.5 nm)/C$_{60}$ (13 nm) and CFB (7 nm)/$\beta$-W(2.5 nm). Hence, the sizable increase in the measured voltage can be attributed to the enhanced SOC of $\beta$-W due to molecular hybridization and additional charge current flowing at the CFB/$\beta$-W interface due to IREE as shown in the Figure \ref{fig:ISHEplot2} (d). In order to understand the enhanced spin-to-charge conversion phenomena further, we also calculated the $\theta_{SH}$ of the heterostructures by using equations \ref{eq:spincurrent} and \ref{eq:ISHE2} \cite{bai2020simultaneously,singh2020large}:
\begin{widetext}
\begin{equation}
    J_{s} = \frac{g_{eff}^{(\uparrow \downarrow)} \gamma^2 h_{rf}^2 \hbar [\gamma 4 \pi M_s + \sqrt{(\gamma 4 \pi M_s)^2 + 4 \omega^2}] }{8 \pi \alpha_{eff}^2 [(\gamma 4 \pi M_s)^2 + 4\omega^2]} \times (\frac{2e}{\hbar}),
    \label{eq:spincurrent}
\end{equation}
\end{widetext}
\begin{widetext}
\begin{equation}
    V_{ISHE} = \frac{w_y L \rho_{NM}}{t_{NM}} \theta_{SH}\lambda_{NM} tanh (\frac{t_{NM}}{2\lambda_{NM}}) J_s,    \label{eq:ISHE2}
\end{equation}
\end{widetext}
where the $\rho_{NM}$ is the resistivity of the $\beta$-W measured by four-probe technique and $L$ is the length of sample. The RF field ($h_{rf}$) and the width of the CPW transmission line ($w_y$) in our measurements are 0.5 Oe (at 15 dBm RF power) and 200 $\mu$m, respectively. The $\lambda_{NM}$ for the $\beta$-W has been taken as $\sim$ 3 nm from the literature \cite{hao2015giant}. Angel dependent ISHE measurements were performed to separate the AMR contribution from the $V_{SYM}$. The contribution of $V_{AMR}$ was found to be one order smaller compared to V$_{ISHE}$. For example, the $V_{AMR}$ and V$_{ISHE}$ for CFW2 heterostructure are found to be $\sim$ 0.15 $\mu$V and $\sim$ 1.25  $\mu$V, respectively (See Supporting Information). The $\rho_{NM}$ for 5 nm $\beta$-W is found to be 250 $\mu \Omega$ cm. Hence, the $\theta_{SH}$ for CFB (7 nm)/$\beta$-W (5 nm) bilayer estimated using equations 6 and 7 is found to be $\sim$ -0.6$\pm$0.01. A similar type of calculation for CFB (7 nm)/$\beta$-W (2.5 nm) bilayer estimates the $\theta_{SH}$ to be $\sim$ -0.67$\pm$0.01. The observed $\theta_{SH}$ value is larger compared to that reported in the literature \cite{sui2017giant,mchugh2020impact,demasius2016enhanced}. The high SOC of our $\beta$-W and higher spin mixing conductance could be responsible for this enhanced $\theta_{SH}$. Further, the interfacial SOC at CFB/$\beta$-W interface can also induce an additive spin-to-charge conversion effect, contributing to the enhancement of $\theta_{SH}$. Such type of interfacial SOC mediated enhanced spin-to-charge conversion has been reported previously for NiFe/Pt and CFB/$\beta$-Ta \cite{allen2015experimental,hao2022significant}. Here, it is important to note that it is difficult to disentangle the IREE and ISHE effect in these type of FM/HM systems. On the other hand, the $g_{eff}^{(\uparrow \downarrow)}$  for CFWC1 and CFWC2 decreases by 70 $\%$ due to the anti-damping phenomena and hence, the reduction in $J_s$ according to equation \ref{eq:spincurrent}. However, the $V_{ISHE}$ for the CFWC1 and CFWC2 are found to be larger than CFW1 and CFW2, respectively (Figure \ref{fig:ISHEplot2}). This leads to the $\theta_{SH}$ value $>$1, calculated using the equation \ref{eq:spincurrent} and \ref{eq:ISHE2} for CFB/$\beta$-W/C$_{60}$ heterostructures. This type of gigantic enhancement of $\theta_{SH}$ cannot be explained by mere bulk ISHE in $\beta$-W. The enhanced $\theta_{SH}$ can be partly attributed to the enhanced bulk SOC of $\beta$-W upon molecular hybridization as predicted by the DFT calculations. Further, our DFT calculations also predict the enhancement of SOC of $\beta$-W layer closer to the CFB/$\beta$-W interface due to the molecular hybridization in the CFB/$\beta$-W/C$_{60}$ heterostructures. The larger interfacial SOC and inversion symmetry breaking at the CFB/$\beta$-W interface makes the scenario favorable for realizing an enhanced interfacial charge current due to the IREE as depicted in Figure \ref{fig:ISHEplot2} (d). Hence, the combination of bulk and interfacial SOC enhancement owing to the strong chemisorption of C$_{60}$ on $\beta$-W can attribute to the sizable increase in the $\theta_{SH}$ in CFB/$\beta$-W/C$_{60}$ heterostructures. 
\par
The enhanced output DC voltage due to the spin pumping upon the C$_{60}$ deposition on $\beta$-W is also consistent with the reduced effective damping value as discussed earlier. The enhanced SOC of $\beta$-W and the structural inversion asymmetry at the CFB/$\beta$-W interface can stabilize the Rashba like states at FM/HM interface \cite{allen2015experimental,hao2022significant}. The IREE mediated spin to charge conversion has received considerable interest after it was discovered at the Ag/Bi interface \cite{sanchez2013spin}. Till the date, most of the IREE effects have been experimentally realized at the all inorganic metal/metal, metal/oxide or oxide/oxide interfaces \cite{koo2020rashba}. Our experiments and theoretical calculations show that the molecular hybridization at the HM/OSC interface can also help in strengthening the Rashba spin-orbit coupling at the FM/HM interface. The Rashba interaction leads to the spin splitting of bands, whose magnitude is dependent on the SOC strength at the interface. Upon the molecular hybridization, the SOC strength of $\beta$-W is further enhanced. This could have lead for a larger Rashba coefficient $\alpha_R$ and hence, a relatively larger IREE at the FM/HM interface. The simultaneous observation of ISHE and IREE by engineering the HM interface with OSC can help in reducing the power consumption of future SOT-MRAM devices. As the CFB/$\beta$-W stacking is employed for fabrication of spin Hall nano oscillators (SHNOs) \cite{zahedinejad2022memristive}, the incorporation organic molecules can also significantly enhance their efficiency. Hence, the HM/C$_{60}$ interface can reduce the power consumption for data storage as well as facilitate in performing efficient spin logic operations.
\section{Conclusion}
In conclusion, we present that a strong interfacial SOC can lead to the larger spin Hall angle in CFB/$\beta$-W bilayer. The thermally evaporated organic C$_{60}$ molecules on CFB/$\beta$-W bilayer leads to a strong chemisorption at the $\beta$-W/C$_{60}$ interface. The experimental and theoretical calculations confirm that the molecular hybridization enhances the bulk as well as interfacial SOC in CFB/$\beta$-W/C$_{60}$ heterostructures.  The strengthening of technologically important SOC manifests an anti-damping phenomena and gigantic $\sim$ 115$\%$ increase in spin-pumping induced output voltage for CFB/$\beta$-W/C$_{60}$ stacking. The control of magnetization dynamics and output efficiency in spintronics devices by the molecular hybridization can be a viable alternative to the other interface engineering and surface alloying techniques. The stabilization of the anti-damping and enhanced spin-to-charge conversion by tuning the bulk as well interfacial SOC via employing the cost effective, abundant organic molecule can pave the way for the fabrication of next generation power efficient spintronics devices.
\section{Acknowledgement}
We acknowledge the Department of Atomic Energy (DAE), the Department of Science and Technology (DST) of the Government of India, and SERB project CRG/2021/001245. A.S. acknowledges the DST-National Postdoctoral Fellowship in Nano Science and Technology. We are also thankful to the Center for Interdisciplinary Sciences, NISER for providing Raman spectroscopy measurement facility.
%\nocite{*}
\bibliography{apssamp}% Produces the bibliography via BibTeX.
\end{document}